\documentclass[a4paper,twocolumn,11pt,unpublished]{quantumarticle}
\pdfoutput=1
\usepackage{amssymb,amsmath,amsfonts}
\usepackage{epsfig,graphicx}
\usepackage{braket}
\usepackage{hyperref}
\usepackage[dvipsnames]{xcolor}
\hypersetup{colorlinks,citecolor=PineGreen,filecolor=black,linkcolor=black,urlcolor=black}
\usepackage[utf8]{inputenc}
\usepackage[english]{babel}
\usepackage[T1]{fontenc}
\usepackage{tikz}
\usepackage{lipsum}
\usepackage[numbers]{natbib}

\begin{document}
\title{Invariant-based control of quantum many-body systems across critical points}



\author{H. Espin{\'o}s}
\affiliation{Department of Physics, Universidad Carlos III de Madrid, Avda. de la Universidad 30, Legan\'es, 28911  Madrid, Spain}
\author{L. M. Cangemi}
\affiliation{Department of Chemistry; Institute of Nanotechnology and Advanced Materials; Center for Quantum
Entanglement Science and Technology, Bar-Ilan University, Ramat-Gan, 52900 Israel}
\author{A. Levy}
\affiliation{Department of Chemistry; Institute of Nanotechnology and Advanced Materials; Center for Quantum
Entanglement Science and Technology, Bar-Ilan University, Ramat-Gan, 52900 Israel}
\author{R. Puebla}
\affiliation{Department of Physics, Universidad Carlos III de Madrid, Avda. de la Universidad 30, Legan\'es, 28911  Madrid, Spain}
\author{E. Torrontegui}
\email{eriktorrontegui@gmail.com}
\affiliation{Department of Physics, Universidad Carlos III de Madrid, Avda. de la Universidad 30, Legan\'es, 28911  Madrid, Spain}

\begin{abstract}
Quantum many-body systems are emerging as key elements in the quest for quantum-based technologies and in the study of fundamental physics. In this study, we address the challenge of achieving fast and high-fidelity evolutions across quantum phase transitions, a crucial requirement for practical applications. We introduce a control technique based on dynamical invariants tailored to ensure adiabatic-like evolution within the lowest-energy subspace of the many-body systems described by the transverse-field Ising and long-range Kitaev models. By tuning the controllable parameter according to analytical control results, we achieve high-fidelity evolutions operating close to the speed limit. Remarkably, our approach leads to the breakdown of Kibble-Zurek scaling laws, offering tunable and significantly improved time scaling behavior. We provide detailed numerical simulations to illustrate our findings, demonstrating scalability with the system size and robustness against noisy controls and disorder, as well as its applicability to a non-integrable system. 
\end{abstract}

\maketitle

\section{Introduction}
Quantum many-body systems are drawing growing interest in modern science, in part, thanks to their great significance in the quest towards scalable quantum-based technological applications, such as in quantum simulation~\cite{Buluta:09,Georgescu:14}, computation~\cite{Nielsen,Albash:18} and metrology~\cite{Giovannetti:04,Giovannetti:11,Toth:14}. A precise understanding of quantum many-body physics has also been proven crucial in other research areas, such as material science and statistical mechanics~\cite{Savary:17,Bruss,Fabrizio}.
These systems exhibit multipartite quantum correlations~\cite{Vidal:03,Amico:08,DeChiara:18} alongside collective and emergent phenomena~\cite{Bruss,Fabrizio}. Due to the limited number of exactly solvable models~\cite{sutherland2004beautiful}, their understanding usually relies on numerical approaches~\cite{schollwock2011density,carleo2017solving}. Moreover, their complexity and exponentially large Hilbert space pose severe limits to their classical simulations \cite{trotzky2012probing}.
For these reasons, controllable quantum many-body systems embody a fascinating testbed to develop and test future quantum applications as well as to explore new physics~\cite{Polkovnikov:11, Eisert:15, Zohar:16}.

Quantum phase transitions (QPTs) appear as a genuine trait of quantum many-body systems~\cite{Sachdev,Vojta:03}, where fluctuations of quantum nature, rather than thermal, may trigger an abrupt change in the properties of a system in its ground state. 
This occurs at a certain critical value of an external and controllable parameter, i.e. at the critical point, where relevant quantities display singular behavior, captured by critical exponents~\cite{Sachdev}. The energy gap typically vanishes at the critical point. 
Hence, QPTs have profound consequences in both in- and out-of-equilibrium dynamics~\cite{Polkovnikov:11,Eisert:15}, such as in defect formation described by the celebrated Kibble-Zurek (KZ) mechanism~\cite{Kibble:76,Zurek:85,Zurek:96,delCampo:14,Zurek:05,Dziarmaga:05,Damski:05,Kibble:07}. Based on simple assumptions, the KZ mechanism successfully predicts a universal scaling relation between the density of excitations, the equilibrium critical exponents of the QPT, and the rate at which the critical point is traversed~\cite{delCampo:14}, even in the presence of decoherence effects~\cite{Dutta:16,Puebla:20}. Such predictions have been experimentally confirmed in a variety of platforms, obeying classical or quantum dynamics~\cite{Chuang:91, Pyka:13,Ulm:13, Lamporesi:13, Hoang:16, Anquez:16,Keesling:19,Qiu:20,Du:23}. The KZ mechanism shows that non-adiabatic excitations are mainly promoted around a QPT, where the adiabatic condition unavoidably breaks, presenting challenges to quantum adiabatic computation or annealing schedules~\cite{Zurek:05,Gardas:18,deGrandi:10b} and quantum state preparation in many-body systems~\cite{Chandra}. Protocols that allow for a fast and robust evolution through QPTs while providing good fidelities with respect to a target are therefore highly valuable. 


Different control techniques have been proposed to speed up an adiabatic evolution, mitigating the loss of adiabaticity caused by a QPT. Ideally, desirable controls i) allow for a fast and very high or unit fidelity evolution, ii) do not require further tunability of the system (i.e. extra and specifically tailored interaction terms that must be controlled), iii) present good scalability with the system size, and iv) are robust against potential imperfections. 
On the one hand, variational quantum controls, such as Krotov~\cite{Krotov,Goerz:19}, machine-learning~\cite{Bukov:18,Huang:22,Brown:23}, gradient-based~\cite{Khaneja:05,Dalgaard:20} and chopped-random basis quantum optimization~\cite{Caneva:11,Doria:11,Power:13,Rach:15,vanFrank:16} protocols. However, the application of these methods to critical systems with many degrees of freedom soon becomes intractable due to the growing number of parameters to be optimized in the cost function. On the other hand, shortcuts-to-adiabaticity protocols~\cite{GueryOdelin:19,Torrontegui:13}, such as counterdiabatic driving (CD)~\cite{delCampo:13} typically grant unit fidelity by construction at the expense of requiring Hamiltonian diagonalization. As an example, CD guarantees a perfect-adiabatic-like evolution for any evolution time, even when a QPT is traversed, as demonstrated in~\cite{delCampo:12,Takahashi:13,Damski:14,delCampo:15}. However, CD demands an added controllable and time-dependent non-local interaction term, which may be troublesome for its experimental implementation. Alternative shortcuts such as those based on invariants of motion \cite{Chen:10} remain challenging due to the lack of dynamical invariants for many-body systems and the complexity of reverse engineering its dynamics. Different protocols have also been proposed, such as local adiabatic~\cite{Roland:02,Richerme:13}, fast-quasiadiabatic (FAQUAD)~\cite{Garaot:15} and minimal-action based protocols~\cite{Kazhybekova:22}. Recent methods based on optimized CD~\cite{Saberi:14,Cepaite:23} significantly enhance both time and fidelity with respect to standard quantum optimal controls, without requiring full knowledge of the system. Nevertheless, these procedures still demand numerical optimization of a growing number of parameters for larger system sizes and further tunability. Hence, these previously proposed protocols do not meet all the conditions simultaneously.

In this article, we propose a control method for quantum many-body systems that satisfies the mentioned requirements while operating close to the quantum speed limit. Its performance is evaluated in the one-dimensional transverse-field Ising model (TFIM)~\cite{Sachdev,Dutta,Vojta:03} and long-range Kitaev (LRK) model~\cite{Kitaev:01,Vodola:14,Alecce:17,Dutta:17,Defenu:19}, resulting in analytical controls. Importantly, 
the protocol preserves the Hamiltonian in its original form, i.e. without requiring extra interaction or control terms, and it is thus amenable to its experimental realization. The method is precisely based on an invariant control~\cite{Torrontegui:14}, applied to the low-energy subspace of the full Hamiltonian. By construction, the dynamics within the low-energy subspace remain perfectly adiabatic at the end of the passage, so that the resulting excitations stem from transitions between ground- and higher-order excited states.  As a consequence, KZ mechanism breaks down: the reduced non-adiabatic excitations no longer depend on the critical exponents and display a much better and tunable scaling with the evolution time. 
The derived control is tested as a function of the system size $N$, yielding similar results in terms of a rescaled evolution time. The obtained results demonstrate a large improvement over standard methods such as a naive linear ramp and FAQUAD (see Appendix \ref{App:A} for details), and corroborate the independence on the universality class, that is, similar results regardless of the critical exponents. Finally, the protocol is shown to be robust against imperfections, such as disorder and noisy controls.

\section{Quantum many-body systems}
We consider two quantum many-body systems where we will illustrate the invariant-based protocol: a TFIM and a LRK~\cite{Sachdev,Dutta,Vojta:03,Kitaev:01,Vodola:14,Alecce:17,Dutta:17,Defenu:19}. Their Hamiltonians can be written as ($\hbar=1$)
\begin{widetext}
\begin{align}\label{eq:TFIM}
    \hat{H}_{\rm TFIM}&=-J\sum_{i=1}^N(g \hat{\sigma}_i^x+ \hat{\sigma}_{i}^z\hat{\sigma}_{i+1}^z),\\
    \label{eq:LRK}
    \hat{H}_{\rm LRK}&=-J\sum_{i=1}^N\left[ \sum_{r>0} (J_r \hat{c}_i^\dagger \hat{c}_{i+r}+d_r \hat{c}_i\hat{c}_{i+r}+{\rm H.c.})-g \hat{n}_i \right],
\end{align}
\end{widetext}
where $\hat{\sigma}_i^{x,y,z}$ denote the spin-$1/2$ matrices at site $i$,  while $\hat{c}_i$ corresponds to fermionic annihilation operator, and $\hat{n}_i=\hat{c}_i^\dagger \hat{c}_i$. In both cases $J$ accounts for an energy scale for the $N$ interacting particles separated by a distance $a$, while $g$ is a dimensionless controllable parameter that uniformly influences all sites. In the context of the TFIM, $g$ denotes the magnetic-field strength, while in the LRK model, it represents the chemical potential. Periodic boundary conditions are taken for both models. 
 These models feature continuous QPTs at a certain critical value $g_c$ of this external parameter $g$. For the TFIM, $g_c=1$. In the LRK, hopping and pairing strengths, $J_r$ and $d_r$, respectively, depend on long-range exponents $\alpha,\beta>1$, that is, $J_r=(N_\alpha r^\alpha)^{-1}$ and $d_r=(N_\beta r^\beta)^{-1}$ with $N_{\gamma}=2\sum_{r=1}^{N/2}r^{-\gamma}$ for $\gamma=\{\alpha,\beta\}$, following Kac's prescription~\cite{Kac:69}. Thus, critical properties of the LRK also depend on $\alpha$ and $\beta$. In the short-range limit, $\alpha,\beta\rightarrow\infty$, the critical point takes place at $g_c=2J$, and the model maps to a TFIM~\cite{Kitaev:01}.

Both Hamiltonians can be decoupled into a collection of independent Landau-Zener systems in momentum space of spinless fermions, which greatly simplifies the dynamics. The full Hamiltonian can be diagonalized as $\hat{H}=\sum_{k}\Psi^{\dagger}_{k}\hat{H}_{k}\Psi_{k}$, with $\Psi^\dagger=(\hat c_k^\dagger,\hat c_{-k})$, where $\hat c_k^\dagger$ is the creation operator of a spinless fermion with momentum $k$. In this momentum space, one can write 
\begin{equation}\label{eq:HTLS}
\hat{H}_{k}=h_{z,k}(g)\frac{\hat{\sigma}_z^k}{2}+h_{x,k}\frac{\hat{\sigma}_x^k}{2},
\end{equation}
such that $\hat{\sigma}_{z}^k=\ket{1}\bra{1}_k-\ket{0}\bra{0}_k$ represents the $z$-component of the Pauli matrices for the quasiparticles of momentum $k=(2n-1)\pi/(Na)$ with $a$ the interparticle distance, and $n\in\{ 1,\ldots,N/2\}$ (see for example~\cite{Sachdev,Vodola:14}). For the TFIM, $h_{z,k}(g)=4J(g-\cos(k a))$, $h_{x,k}=4J \sin(ka)$, while for the LRK, $h_{z,k}(g)=g-4J\sum_{r>0}J_r\cos(k r a)$ and $h_{x,k}=-2J \sum_{r>0}d_r\sin(k r a)$. The minimum energy gap takes place at the QPT, at $g=g_c$. In the thermodynamic limit, $N\rightarrow \infty$, one finds $\Delta(g)\propto |g-g_c|^{z\nu}$ with $z\nu=1$ in the TFIM, while $z$ and $\nu$ depend on the range of the interactions in the LRK~\cite{Defenu:19}. 

Our goal consists in devising controls $g\equiv g(t)$ that drive an initial ground state in one phase,  $\ket{\psi(t=0)}=\ket{\phi_0(g_0)}$ with $g_0>g_c$, into the ground state in the other phase across the QPT ($g_1<g_c)$, namely, $\ket{\psi(\tau)}=\ket{\phi_0(g_1)}$ where $\ket{\psi(\tau)}=\mathcal{U}(\tau,0)\ket{\psi(0)}$ and $\mathcal{U}(t_1,t_0)=\mathcal{T}e^{-i\int_{t_0}^{t_1}dt'H(t')}$ is the time-evolution operator. Since $\langle\phi_0(g_0) | \phi_0(g_1) \rangle\approx 0$, the minimum time to perform this evolution, known as quantum speed limit (QSL), amounts to $\tau_{\rm QSL}=\pi/\Delta$~\cite{Hegerfeldt:13}, with $\Delta$ the minimum energy difference between the ground and first excited states of the complete system.  



The formalism of dynamical invariants~\cite{LR} can be readily applied to the TFIM and LRK models to control the lowest-energy subspace with momentum $k_0=\pi/(Na)$. 
Indeed, the control $g(t)$ is analytically determined from the equation
\begin{equation}
    h_{z,k_0}(g)=\frac{\ddot f_z+f_z h_{x,k_0}^2}{h_{x,k_0}\sqrt{K-f_z^2-\left(\dot f_z/h_{x,k_0}\right)^2}},
\end{equation}
where $f_z\equiv f_z(t)$ and $K$ are an arbitrary function of time and an arbitrary constant satisfying boundary conditions at $t_B=0,\tau$ (see Appendix \ref{App:B}). It is noteworthy that even though $g(t)$ is derived from the lowest-energy momentum subspace, which exhibits maximal non-locality in the original non-diagonal frame, its application to the complete Hamiltonians \eqref{eq:TFIM} and \eqref{eq:LRK} only entails local interactions that affect each site individually.
As an illustration, the control $g(t)$ for a TFIM with $N=200$ spins employing a fifth-order polynomial for $f_z$ is plotted in Fig.~\ref{fig1}(a). The minimum evolution time for this choice results in $\tau\gtrsim 1.2\tau_{\rm QSL}$ (see Appendix \ref{App:B.2}).


\begin{figure*}[t!]
    \centering
    \includegraphics[width=1\linewidth]{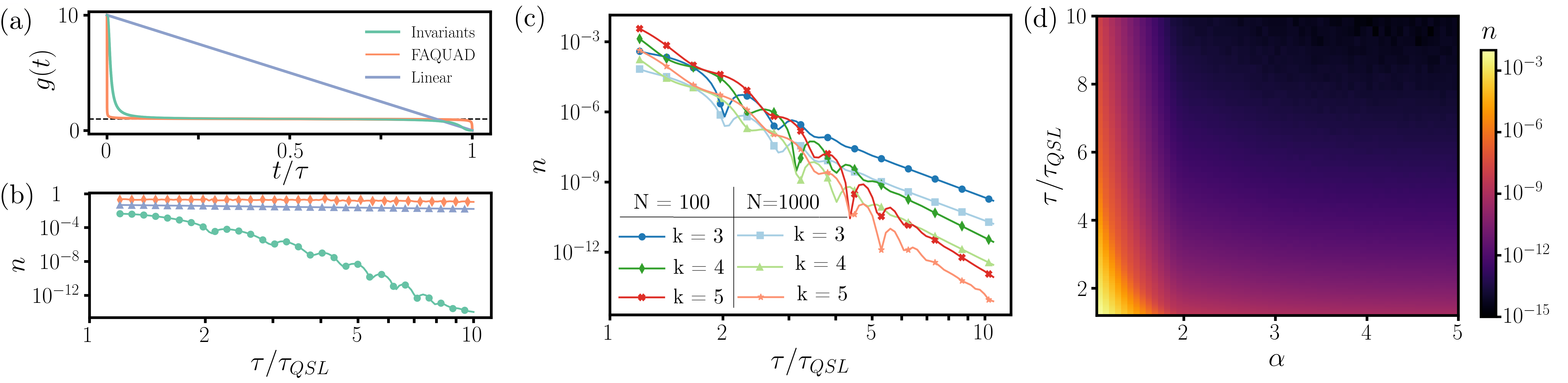}
    \caption{(a) Sketch of the invariant, FAQUAD, and linear controls starting at $g(0)=10$ and ending at $g(\tau)=0$, simulated for $N=200$ spins. (b) Scaling of the density of excitations (log-log) for the TFIM and fixed $N=200$ spins for the three protocols. (c) TFIM invariant for $N=100$ and $N=1000$ as a function of the rescaled quench time $\tau/\tau_{\rm QSL}$, including also the $\tau^{-8}$ and $\tau^{-10}$ protocols. (d) Scaling of $n$ in the LRK for $N=100$ fermions, $\beta\rightarrow\infty$, as a function of the quench time $\tau$ (in units of $\tau_{\rm QSL}$) and the long-range exponent  $\alpha$.  }
    \label{fig1}
\end{figure*}

\section{Breakdown of Kibble-Zurek mechanism and quasiadiabatic dynamics}
Critical features of the QPT typically leave their footprint on the dynamics. The KZ mechanism~\cite{delCampo:14,Zurek:05} predicts an unavoidable departure from an adiabatic evolution due to a vanishing energy gap $\Delta(g)\propto |g-g_c|^{z\nu}$. 
This adiabatic-impulse approximation in the vicinity of the QPT leads to the celebrated KZ scaling relations, such as the density of excitations $n\sim \tau^{-d\nu/(z\nu+1)}$ for a linear ramp across the QPT with $\dot{g}\propto 1/\tau$, and $d$ being the spatial dimension of the system~\cite{Polkovnikov:11,Zurek:05,delCampo:14,deGrandi:10b}. Here, the density of excitations is defined as
\begin{equation}
    n=1-\frac{2}{N}\sum_{k>0}|\langle \phi_{k,0}|\psi_k(\tau)\rangle|^2,
\end{equation}
with $\ket{\phi_{k,0}}$ the ground state at $g(\tau)$ in the $k$ subspace, and $\ket{\psi_k(\tau)}$ the evolved state. 
However, an engineered control $g(t)$ that strictly ensures an adiabatic-like evolution within the lowest-energy subspace will not follow KZ mechanism. Under this $g(t)$,  the system evolves as being gapped throughout the whole quench, i.e. as if it had no critical or singular traits\footnote{Assuming that the gap only closes between ground and first excited state, as for short-ranged interactions. For fully-connected models, however, this is no longer true as all eigenstates may coalesce at the critical point.}. This results in the breakdown of the KZ mechanism, i.e. $n\sim \tau^{-dv/(z\nu+1)}$ will no longer be applicable. 


The invariant-based control $g(t)$ ensures adiabaticity at the end of the passage in the lowest-energy subspace. The dynamics within higher-order subspaces may be quasi-adiabatic due to their larger energy separation. The resulting time scaling, different from KZ prediction, can be obtained by relying on a simple Landau-Zener system. 
Its Hamiltonian, with minimum energy gap $\Delta_1$, is quenched from its ground state following a control $g(t)$. 
The control $g(t)$ ensures the target adiabatic state transfer in a subspace with gap $\Delta_0<\Delta_1$. Applying adiabatic perturbation theory~\cite{deGrandi:10}, one can show that the density of excitations $n$ in the subspace with larger gap scales as $n\propto |\dot{g}(t_B)|^2$, and similarly for the infidelity $\mathcal{I}=1-\mathcal{F}$ with $\mathcal{F}=\lvert\langle\phi_0(g_1)|\psi(\tau)\rangle|^2$ (see Appendix \ref{App:B.1}). For linear ramps, $\dot{g}(t_B)\propto \tau^{-1}$ resulting in the standard $\tau^{-2}$ time scaling for quasi-adiabatic dynamics. In our case, however, $\dot{g}(t)$ depends on the auxiliary function $f_z$. Choosing a polynomial, $\dot{g}(t_B)\propto \dddot{f_z}(t_B)$ since $\dot{f}_z(t_B)=\ddot{f}_z(t_B)=0$ by construction (see Appendix \ref{App:B}, as well as $\dddot{f}_z(t_B)\propto \tau^{-3}$, we have $n\sim \tau^{-6}$. This scaling can be further improved by noticing that the auxiliary function $f_z$ is arbitrary, and can be constructed requiring all first $k-1$ time-derivatives of $f_z$ to vanish at the boundaries $t_B$. In this manner, one finds 
\begin{equation}
    n\propto \Biggl\lvert\frac{d^{k}f_z}{dt^{k}}(t_B)\Biggr\rvert^2 \propto \tau^{-2k},
\end{equation}
for $k\geq 3$, so that the time-scaling across QPTs can be arbitrarily enhanced. 

\section{Numerical results}
These theoretical predictions are corroborated by numerical simulations in the TFIM and LRK for different long-range exponents. For completeness, we compare the invariant-based control with a linear ramp and FAQUAD designed for the low-energy subspace. The results for the density of excitations $n$ for a TFIM with  $N=200$ spins, $g(0)=10$ and $g(\tau)=0$, are plotted in Fig.~\ref{fig1}(b), which reveal an improvement of the invariant control by several orders of magnitude with respect to these other protocols, and clearly breaking the KZ scaling law. Note that already for $\tau\approx 2\tau_{\rm QSL}$, we obtain $n\sim 10^{-5}$, improving by 4 orders of magnitude at $\tau\approx 5\tau_{\rm QSL}$. In addition, the time scaling agrees with the $\tau^{-2k}$ law independently on the system's size, as shown in Fig.~\ref{fig1}(c) for $k=3$, $4$ and $5$ with suitably chosen controls (see Appendices \ref{App:C.1} and \ref{App:C.2}).  Similar results are obtained for the infidelity. The resulting dynamics are agnostic to the critical features,  as shown for the LRK with different long-range exponent (cf. Fig.~\ref{fig1}(d)). However, extremely long-ranged interactions may cause gapless higher-excited states at the critical point, such as in mean-field models. This makes the derived control unfit to provide speedup adiabatic evolutions for $\alpha\rightarrow 1$ in the LRK.

\subsection{Robustness} Having shown that the invariant-based control fulfills the requirements i), ii), and iii) mentioned in the introduction, namely, high fidelities for fast evolutions, minimal tunability, and good scalability with the system size, we bring our attention to the iv) criterion.  That is, its robustness against potential imperfections. For that, we analyze the performance of the invariant-based protocol designed for an ideal TFIM under noisy control as well as under random disorder. 
First, we consider a $\hat{H}_{\rm TFIM}$ with $g(t)\rightarrow g(t)+\eta(t)$ (cf. Eq.~\eqref{eq:TFIM}), being $\eta(t)$ a white Gaussian noise that accounts for fluctuations on the control, so that $\langle \eta(t)\rangle=0$ and $\langle \eta(t)\eta(t')\rangle=W^2$, with $W$ the strength of the fluctuations. The resulting approximate master equation adopts a Lindblad form for each subspace~\cite{Dutta:16}
\begin{equation}
\dot{\hat{\rho}}_k=-i[\hat{H}_{0,k}(t),\hat{\rho}_k(t)]+\mathcal{D}_{1,k}[\hat{\rho}_k(t)].
\end{equation}
The incoherent part is given by 
\begin{equation}
    \mathcal{D}_{1,k}[\hat{\rho}_k]=\Gamma (\hat{L}_k\hat{\rho}_k \hat{L}_k^\dagger-\frac{1}{2}\{\hat{L}_k^\dagger \hat{L}_k,\hat{\rho}_k \}),
\end{equation}
with $\Gamma=4J^2W^2$ and $\hat{L}_k=\hat{\sigma}_z^k$ (see Appendix \ref{App:C.4}).

\begin{figure}[t]
    \centering
    \includegraphics[width=1\linewidth]{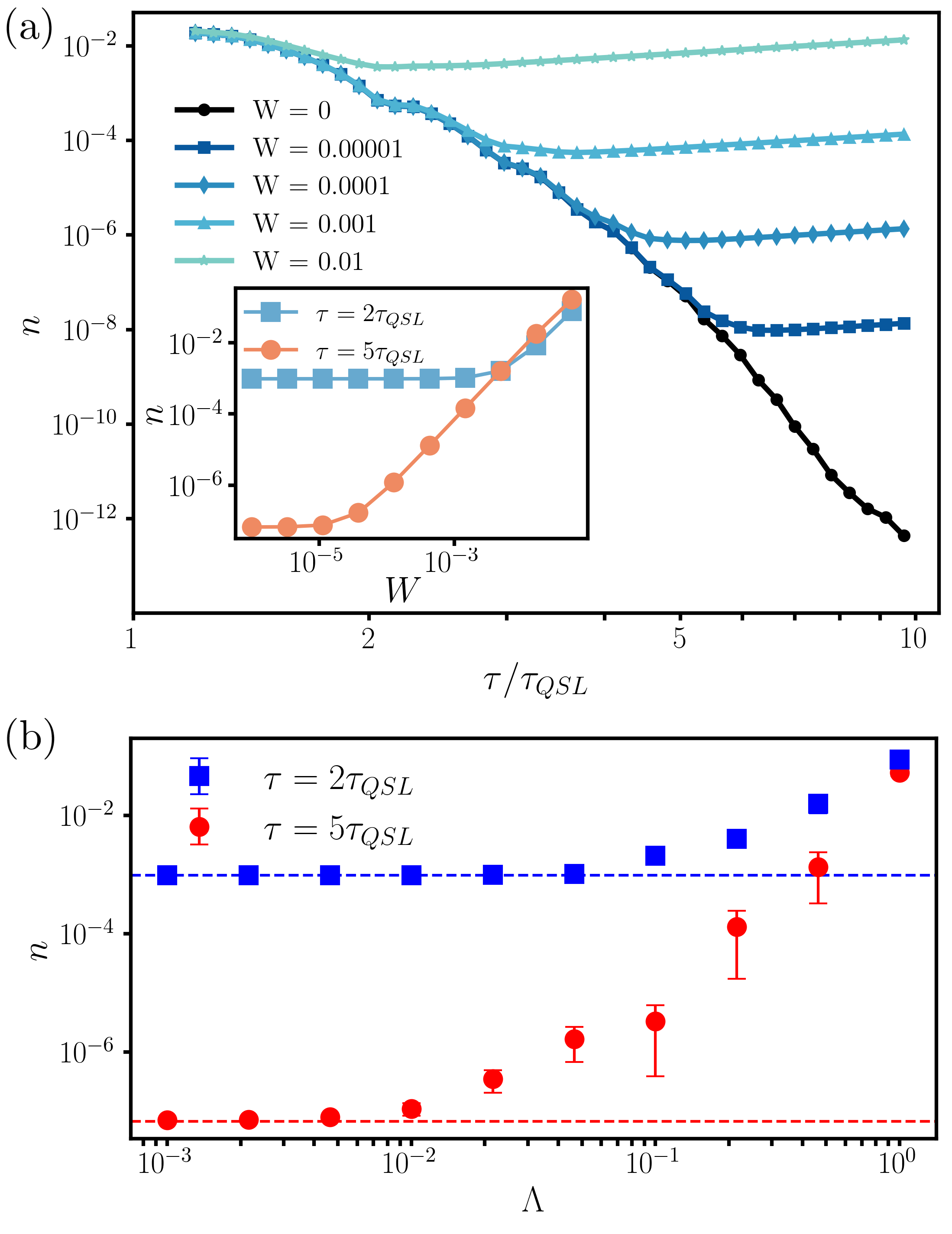}
    \caption{Robustness for noisy control. (a) The density of excitations as a function of the rescaled quench time, for $N=50$ spins and for different white Gaussian noise strengths $W$ affecting the control. (b) Impact of local random disorder $\lambda_i$ in the average density of excitations (after $10$ repetitions) as a function of the disorder strength $\Lambda$ for $N=50$ spins, $g(0)=10$ and $g(\tau)=0$. Error bars correspond to a standard deviation, while dashed lines represent the ideal $\Lambda=0$ case. See main text for details.  }
    \label{fig2}
\end{figure}

The results are gathered in Fig.~\ref{fig2}(a), where we show the density of excitations $n$ as a function of the quench time $\tau$ for $N=50$ spins for different noise strengths $W$. As it can be seen, the invariant-based control performs well even for moderate values of $W$, still revealing its characteristic $\tau^{-2k}$ scaling with $k=3$. Its performance is better illustrated in the inset of Fig.~\ref{fig2}(a), where $n$ is plotted against $W$ for two quench times, namely, $\tau/\tau_{\rm QSL}=2$ and $\tau/\tau_{\rm QSL}=5$. Similar results are obtained for larger system sizes. Recall that the critical exponents are irrelevant for the scaling properties, and thus no universal anti-KZ scaling can be found under this protocol (see however Refs.~\cite{Dutta:16,Puebla:20}). 


We also examine the impact of interaction-strength disorder. We consider dimensionless, independent and identically distributed uniform random variables $\lambda_i\in [1-\Lambda,1+\Lambda]$, so that~\cite{Young:96,Caneva:07,Sadhukan:20}
\begin{equation}
    \hat{H}_{\rm TFIM}=-J \left(g\sum_{i=1}^N\hat{\sigma}_i^x+\sum_{i=1}^N\lambda_i \hat{\sigma}_i^z\hat{\sigma}_{i+1}^z\right).
\end{equation}
The ideal case considered above is recovered for $\Lambda=0$ (cf. Fig.~\ref{fig1} and Eq.~\eqref{eq:TFIM}). The dynamics are solved for different disorder realizations $\{\lambda_i\}$, and then the obtained results are averaged over such realizations (see Appendix \ref{App:C.4} for details). The averaged density of excitations as a function of $\Lambda$ is plotted in Fig.~\ref{fig2}(b) for $N=50$ spins. The ideal result is achieved for $\Lambda = 0$, while even for a non-negligible disorder, e.g. $\Lambda\approx 0.2$, and close to $\tau_{\rm QSL}$ the density of excitations can be found below $10^{-3}$. As for the noisy control, the invariant-based control designed for the ideal TFIM results in robust and highly adiabatic-like evolutions operating close to the quantum speed limit.


\subsection{Application to a non-integrable system}
\begin{figure}[t!]
\begin{center}
\includegraphics[width=1\linewidth]{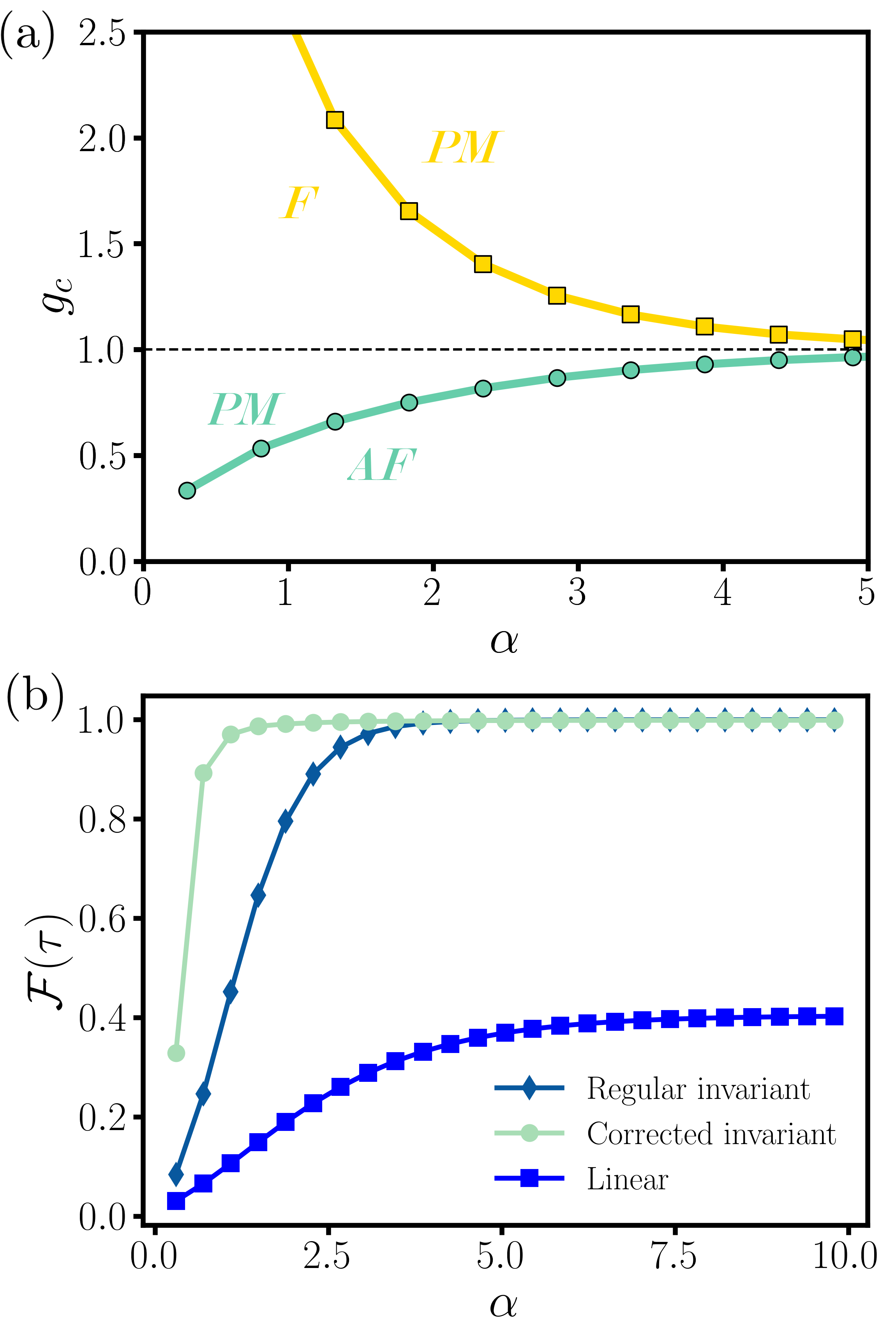}
\caption{(a) Phase diagram ($g-\alpha$) of the long-range transverse field Ising model: The green circles and yellow squares represent the critical magnetic field $g_c$ for antiferromagnetic and ferromagnetic couplings, respectively, delineating the boundaries between the paramagnetic phase (PM) and the antiferromagnetic (AF) and ferromagnetic (F) phases. The dashed line illustrates the critical magnetic field $g_c$ for the nearest-neighbor case ($\alpha\rightarrow\infty$). (b) Fidelity of the regular invariant protocol, the invariant protocol corrected by the introduction of a parameter to take into account the critical magnetic field strength shift, and the linear protocols as a function of the exponent of the long-range interactions $\alpha$. The simulation is performed for $N=12$ spins, with $g(0)=10$, $g(\tau)=0.01$ and $\tau = 4\tau_{\rm QSL}$.}
\label{LRTFIM_supp}
\end{center}
\end{figure}

The previously discussed Hamiltonians lend themselves to representation as a collection of two-level systems. A slightly more encompassing non-integrable Hamiltonian is the transverse-field Ising model including long-range interactions. This system is described by the Hamiltonian
\begin{equation}
    \hat{H}=-J\left(\sum_{i=1}^Ng \hat{\sigma}_i^x+(-1)^p \sum_{i=1}^N\sum_{j>i}^N\frac{1}{\lvert i-j\rvert^\alpha}\hat{\sigma}_{i}^z\hat{\sigma}_{j}^z\right),
\end{equation}
where $\alpha \geq 0$ characterizes the decay behavior of the long-range interactions, while the parameter $p = \{0,1\}$ designates whether the interaction is ferromagnetic ($p=1$) or antiferromagnetic ($p=0$). As shown in Ref.~\cite{Puebla:19} and reproduced here in Fig.~\ref{LRTFIM_supp}(a), the introduction of long-range interactions induces a shift in the critical magnetic field $g_c$ with respect to the TFIM~\eqref{eq:TFIM}. In ferromagnetic systems, these interactions lead to a larger critical magnetic field $g_c$, whereas in antiferromagnetic systems, they reduce the magnitude of the critical point compared to $g_c=1$ in~\eqref{eq:TFIM}. 

We focus our analysis on the system governed by antiferromagnetic interactions. As it was shown in Fig.\ref{fig1}(a), the invariant approach is characterized by a rapid variation of the magnetic strength when far from the critical point and a smooth transition during its traversal. Due to the shift in the value of the critical magnetic field strength, we now anticipate to encounter challenges when considering systems with long-range interactions, since the smooth part of the traversal is not performed at the right value of magnetic field strength. To accommodate this shift, we introduce a parameter into the reference TFIM Hamiltonian from which the invariant is derived,
\begin{equation}\label{eq:HnonInt}
    \hat{H}_{\rm ref}=-J\sum_{i=1}^N(g \hat{\sigma}_i^x+\lambda\hat{\sigma}_{i}^z\hat{\sigma}_{i+1}^z),
\end{equation}
where $\lambda$ serves as the dimensionless parameter introduced for convenience. Despite this modification, the system remains amenable to diagonalization in the form of Eq.~\eqref{eq:HTLS}, with adjusted terms $h_k^z(g) = 4J(g-\lambda\cos (k a))$ and $h_k^x = 4J\lambda\sin (k a)$. 

In the thermodynamic limit ($k_0=\pi/N \rightarrow 0$), the QPT occurs at $g_c=\lambda$. Leveraging Hamiltonian Eq.~\eqref{eq:HnonInt}, we can design the invariant protocol by intentionally selecting $\lambda$ to match the critical magnetic field identified numerically for the non-integrable long-range transverse field Ising model~\cite{Puebla:19}.

In Fig.~\ref{LRTFIM_supp}(b) we present the fidelity after the transition through the quantum critical point for a system of 12 spins going from $g(0)=10$ to $g (\tau)=0.01$ in $\tau = 4\tau_{\rm QSL}$. We observe that the original protocol based on invariants for the nearest-neighbor case outperforms the linear protocol across all values of $\alpha$, showcasing its superior robustness. Moreover, with the introduced modification in the control, the protocol's efficacy extends to longer-range interactions, surpassing $\alpha=2$. As we approach $\alpha\rightarrow0$, it is expected that all protocols fail, as the system exhibits diverging all-to-all interactions. For $\alpha=5$, both protocols based on invariants (original and modified) achieve an infidelity of $10^{-3}$.

\section{Conclusions} In this work we have obtained families of protocols to control quantum many-body systems across QPTs providing high-fidelity evolutions and operating close to the quantum speed limit. The method is based on an invariant control of the low-energy subspace of the many-body system. Our proposed protocol does not require the addition of extra terms in the Hamiltonian, as is the case for counterdiabatic driving, and just requires a particular time-dependent shape of the controllable external parameter, thus making it amenable for its experimental realization. This is illustrated in two relevant models (TFIM and LRK), where we work out analytical expressions for the time-dependent control for any system size. By construction, the resulting dynamics are agnostic to the critical features of the QPT, and therefore, KZ mechanism breaks down, leading to much better time scalings. The robustness of the invariant-based method is studied against two distinct sources of imperfections, noisy control, and disorder in the interaction strengths. Finally, the method is applied to a non-integrable system, showcasing its potential applicability to more complex scenarios. The proposed method can be readily applied to current experimental setups, with application in quantum adiabatic computation and state preparation. 

\acknowledgments 
The authors are grateful to A. del Campo for his comments. We acknowledge financial support from the Spanish Government via the projects PID2021-126694NA-C22 (MCIU/AEI/FEDER, EU) and TSI-069100-2023-8 (Perte Chip-NextGenerationEU). H. E. acknowledges the Spanish
Ministry of Science, Innovation and Universities for funding through the FPU program (FPU20/03409). E. T. acknowledges the Ram{\'o}n y Cajal (RYC2020-030060-I) research fellowship. A. L. acknowledges support from the Israel Science
Foundation (Grant No. 1364/21)

\onecolumn\newpage
\appendix

\section{Fast quasi-adiabatic dynamics}\label{App:A}

We are interested in the control of the Hamiltonian
\begin{align}
    \label{ham}
    \hat H(t)=h_x\frac{\hat\sigma_x}{2}+h_z(t)\frac{\hat\sigma_z}{2},
\end{align}
such that the initial ground state $|-(0)\rangle$ of $\hat H(0)$ with $h_z(0)=h_z^0$ is evolved in a time $\tau$ to the target final state that corresponds to the ground state $|-(\tau)\rangle$ of $\hat H(\tau)$ with  $h_z(\tau)=h_z^F$. Note that the control $h_x(t)=h_x$ is time independent  and $h_y(t)=0$ $\forall t$. The instantaneous eigenvectors of the Hamiltonian are given by
\begin{align}
    \label{eigenS}
    \ket{+(t)}= \cos{\frac{\theta}{2}}\ket{0}+\sin{\frac{\theta}{2}}\ket{1},\nonumber\\
    \ket{-(t)}= \sin{\frac{\theta}{2}}\ket{0}-\cos{\frac{\theta}{2}}\ket{1},
\end{align}
with $\ket{0}$ and $\ket{1}$ being the eigenstates of $\hat\sigma_z$ and $\tan{\theta}= h_x/h_z(t)$. The corresponding eigenenergies $E_\pm(t)=\pm\frac{1}{2}\sqrt{h_x^2+h_z(t)^2}$ are depicted in Fig. \ref{shapes}(a) showing an avoided crossing $\Delta E=E_+-E_-$ with a minimum gap $\Delta E_{\rm min}=h_x$ at $h_z(t)=0$. The Hamiltonian exhibits a typical Landau-Zener passage when the rate of change of the gap $\Delta E$ is slow compared to the intrinsic dynamics of the system 
\begin{equation}
    \label{adia}
     \mu(t)=\bigg|\frac{\braket{-(t)|\partial_t+(t)}}{E_-(t)-E_+(t)}\bigg|=\bigg|\frac{\bra{-(t)}\frac{\partial\hat H}{\partial t}\ket{+(t)}}{[E_-(t)-E_+(t)]^2}\bigg|=\bigg|\frac{h_x\dot h_z(t)}{2[h_x^2+h_z^2(t)]^{3/2}}\bigg|\ll 1\quad\forall t .
\end{equation}
In case of such adiabatic evolution, the state of the system follows the instantaneous eigenstates given by Eq.~\eqref{eigenS} up to a global phase. Equation~\eqref{adia} is the condition for an adiabatic driving at the expense of a global constraint on the control change rate, during the whole process, to its minimum value, $\Delta E_{\rm min}$, leading to slow passages. 

The fast quasi-adiabatic dynamics formalism \cite{Garaot:15} aims at preserving the adiabatic condition locally in time. This can be done by making the adiabaticity parameter constant $\mu(t)=c_\mu\ll 1$ such that the transition probability is equally delocalized during the whole process. Integrating $\mu(t)=c_\mu$ with the initial and final conditions $h_z(0)=h_z^0$ and $h_z(\tau)=h_z^F$ sets a value for the integration constant and the adiabaticity parameter $c_\mu$ leading to a lengthy but analytical control. Its shape is depicted in Fig. \ref{shapes}(b), as shown,  the derived $h_z(t)$ allows fast variations when the avoided crossing opens and remains constant in the proximity to $\Delta E_{\rm min}$.

\section{Invariant-based control}\label{App:B}
In the following, we discuss how to construct a time-dependent protocol based on a invariant control of a two-level system, which will later be identified with the lowest-energy subspace of the full system. For that, we start from the most general two-level Hamiltonian, $\hat H(t)=\frac{1}{2}\sum_{a} h_a(t)\hat\sigma_a$ with $[\hat\sigma_b,\hat\sigma_c]=2i\epsilon_{abc}\hat\sigma_a$ where $\epsilon_{abc}$ is the antisymmetric Levi-Civita tensor, with $a,b,c\in\{x,y,z\}$. 
Associated with this Hamiltonian there are time-dependent Hermitian invariants of motion in the Schr\"{o}dinger picture $\hat I(t )$ satisfying 
\begin{equation}
    \label{def}
    \frac{d\hat I}{dt}\equiv\frac{\partial \hat I(t)}{\partial t}+i[\hat H(t),\hat I(t)]=0.
\end{equation}
A dynamical wave function $|\Psi(t)\rangle$ which evolves 
with $\hat H(t)$ 
can be expressed as a linear combination of invariant modes
\begin{align}
    \label{wave}
    |\psi(t)\rangle=\sum_{j}c_je^{i\alpha_j}|\phi_j(t)\rangle,
\end{align}
where the $c_j$ are constant expansion coefficients, $\alpha_j$ are the Lewis-Riesenfeld phases, fulfilling
\begin{align}
    \frac{d\alpha_j}{dt}=\langle\phi_j(t)|i\frac{\partial}{\partial t}-\hat H(t)|\phi_j(t)\rangle,
\end{align}
and the invariant eigenvectors of $\hat I(t)$,  $|\phi_j(t)\rangle$ form a complete set that satisfies 
\begin{align}
    \hat I(t)|\phi_j(t)\rangle=\lambda_j|\phi_j(t)\rangle,
\end{align}
where $\lambda_j$ are the constant eigenvalues. Note that the previous expansion sets the dynamical wave function to evolve along the instantaneous eigenstates of $\hat I(t)$, while excited states of $\hat H(t)$ are allowed to be populated during the dynamics, in contrast to an adiabatic passage. 

 %
\begin{figure}[t!]
\begin{center}
\includegraphics[width=17cm]{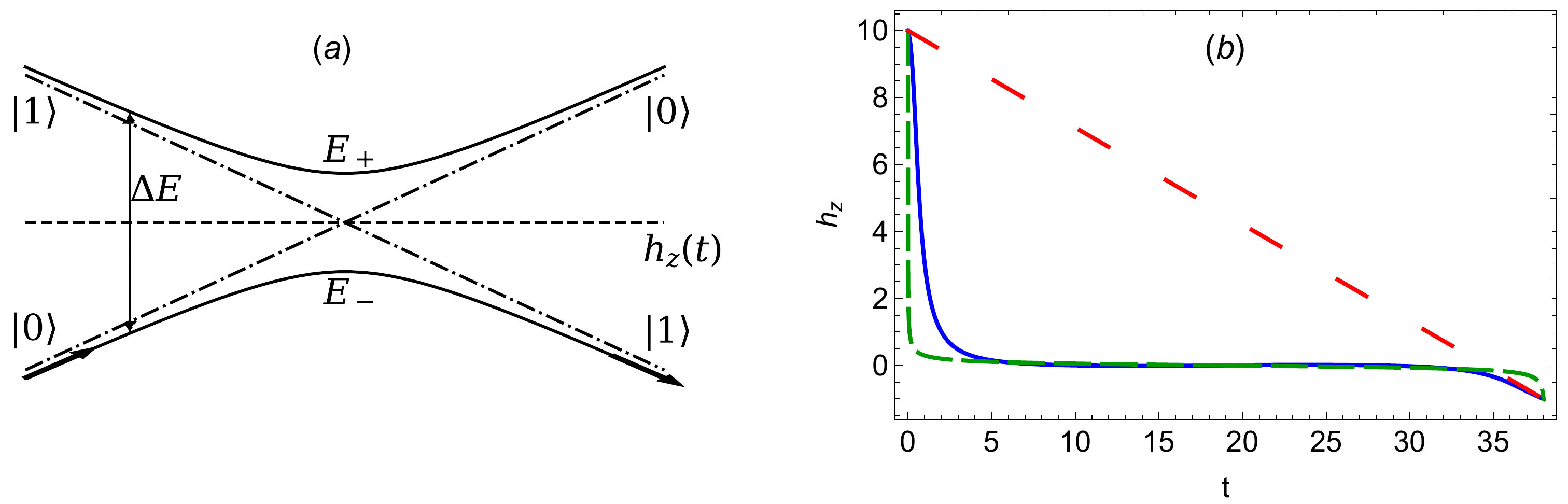}
\caption{(a) Landau-Zener passage through the avoided crossing of Hamiltonian~\eqref{ham}. (b) Different shapes of the $h_z$ driving, invariant-based (blue-solid
line), FAQUAD (green-dashed line), and linear (red-dotted line) as a function of $t$ for the minimum ramp time $\tau=38$ of Fig. \ref{fid}. Parameter values: $h_z(0)=10$, $h_z(\tau)=-1$, and $h_x=0.1$.
 }
\label{shapes}
\end{center}
\end{figure}

If the invariant is a member of the dynamical algebra, it can be written as
\begin{align}
    \label{inv}
    \hat I(t)=f_x(t)\frac{\hat \sigma_x}{2}+f_y(t)\frac{\hat \sigma_y}{2}+f_z(t)\frac{\hat \sigma_z}{2},
\end{align}
where $f_a(t)$ are real, time-dependent functions. 
Replacing  Eqs. (\ref{ham}) and (\ref{inv})
into Eq. (\ref{def}), the  functions $h_a(t)$ and $f_a(t)$ satisfy 
\begin{align}
\label{sisSU2}
\left(\begin{array}{c} 
\dot f_x \\
\dot f_y   \\
\dot f_z 
\end{array} \right)=
\left(\begin{array}{ccc} 
0&   f_z & -f_y  \\
-f_z &   0& f_x \\
f_y      &  -f_x & 0 
\end{array} \right)
\left(\begin{array}{c} 
h_x  \\
0\\
h_z
\end{array} \right).
\end{align}

Usually, these coupled equations are interpreted as a linear system of ordinary differential 
equations for $f_a(t)$ when the $h_a(t)$ components of the Hamiltonian are known. 
However, it is possible to adopt the opposite perspective and consider them as an algebraic system to be solved for 
the $h_a(t)$, when the $f_a(t)$ are given.
When inverse engineering the dynamics, the Hamiltonian is usually given at initial and final times. In the systems that we analyse in this paper, $h_z(0)=h_z^0$ and  $h_z(\tau)=h_z^F$ at the final time 
 $\tau$, while during the whole evolution  $h_x(t)=h_x=const$ and $h_y(t)=0$.  
In general, the invariant $\hat I$ (or equivalently the $f_a(t)$) is chosen   
to drive,  through its eigenvectors,  the initial states of the Hamiltonian $\hat H(0)$ to the states of the final $\hat H(\tau)$. 
This is ensured by imposing at the boundary times $t_B=0, \tau$,  the ``frictionless conditions'' $[\hat H(t_B),\hat I(t_B)]=0$,
\begin{align}
    \label{fric}
    f_a(t_B)h_b(t_B)-f_b(t_B)h_a(t_B)=0,\;\;  a\neq b,
\end{align}
such that the initial and final states become eigenstates of both $\hat H$ and $\hat I$. 
The system \eqref{sisSU2} can be inverted if 
\begin{align}
    \label{gamma}
    f_x^2+f_y^2+f_z^2=K,
\end{align}
with $K$ an arbitrary constant, yielding to an infinite family of solutions ($f_z$ is arbitrary)
\begin{align}
\label{control}
    h_z(t)=\frac{\ddot f_z+f_zh_x^2}{\sqrt{K-f_z^2-\frac{\dot f_z^2}{h_x^2}}}\frac{1}{h_x},
\end{align}
where we made use of the fact that $\dot h_x(t)=0$. The Hamiltonian engineering strategy is complete by interpolating $f_z(t)$ with a simple polynomial or following some more sophisticated approach to satisfy the six boundary conditions ~\eqref{fric} at $t_B$ that explicitly take the form 
\begin{align}
    \label{co1}
    f_z(t_B)=h_z(t_B)\sqrt{\frac{K}{h_x^2(t_B)+h_z^2(t_B)}},\quad \dot f_z(t_B)=0, \quad\ddot f_z(t_B)=0,
\end{align}
and solve for the control $h_z(t)$ in Eq.~\eqref{control}.

%
%
%
%
\begin{figure}[t!]
\begin{center}
\includegraphics[width=17cm]{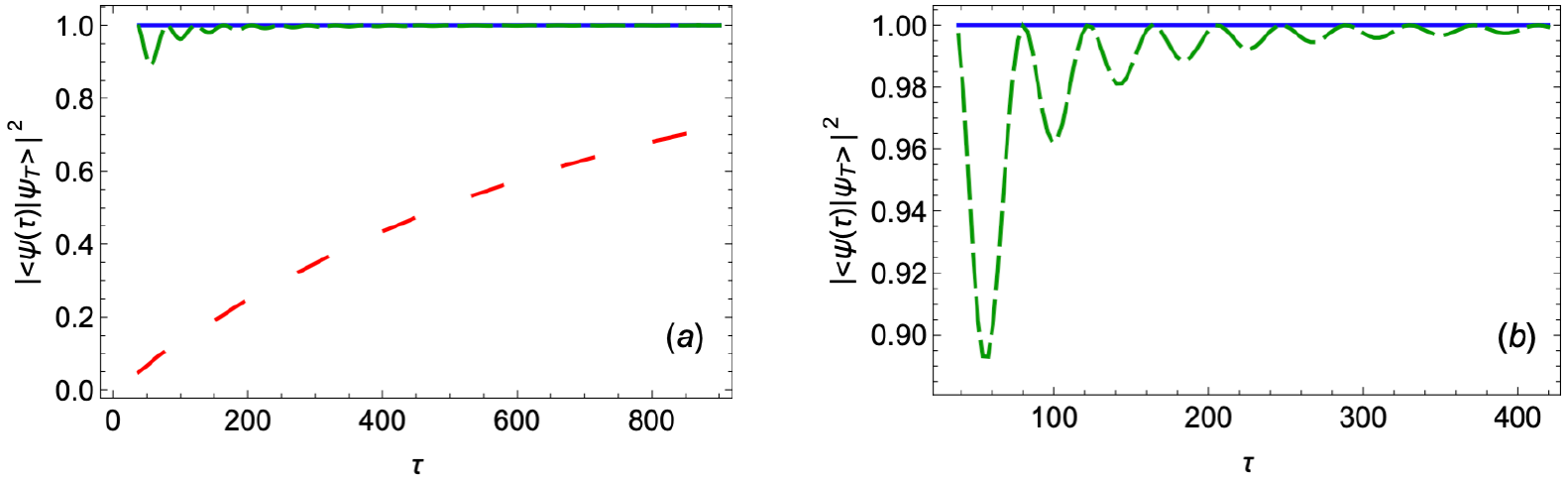}
\caption{(a) Fidelity as a function of the passage time $\tau$ for three different control strategies, invariant-based (blue-solid line), FAQUAD (green-dashed line), and linear (red-dotted line). (b) Close-up for short times. 
Parameter values as in Fig.~\ref{shapes}.
 }
\label{fid}
\end{center}
\end{figure}
%
%

In Fig. \ref{fid}(a)-(b) we plot the fidelity $\mathcal{F}=|\langle \psi(\tau)|\psi_T\rangle|^2$ of the final evolved state with the target final ground state $|\psi_T\rangle=|-(\tau)\rangle$ for three different drivings. This includes the invariant-based reverse engineering, fast quasi-adiabatic (FAQUAD)~\cite{Garaot:15}, and linear drivings of the control $h_z(t)$ as a function of the duration time $\tau$. The corresponding controls are depicted in Fig. \ref{shapes}(b) for the minimum $\tau$. A simple linear ramp leads to a very bad fidelity and only in extremely slow processes is able to reach the desired state $|-(\tau\rangle)$. A more sophisticated protocol, the FAQUAD,  delocalizes the transition probability along the whole dynamical process and allows it to reach $|-(\tau)\rangle$ at shorter times. However, when the process is no longer adiabatic, this process also fails, leading to a lessening of the fidelity. 
 On the contrary, the invariant-based formalism leads to $h_z(t)$ protocol that drives the system to the final target state with unit fidelity,  regardless of the evolved time $\tau$. In order to  find solutions with real functions the condition 
\begin{align}
    \label{condition}
    f_z^2+\frac{\dot f_z^2}{h_x^2}\leq K\quad\forall t 
\end{align}
must be satisfied, see 
Eq.~(\ref{control}). This sets a minimum final time $\tau_{\rm min}$ that 
depends on the ansatz to interpolate $f_z(t)$. 

\subsection{Adiabatic perturbation theory}\label{App:B.1}
In the limit of slow parametric changes, we can find an approximate solution of the Scrhödinger equation,
\begin{equation}\label{SMeq:schro}
i\partial_t\ket{\psi}=\hat{H}\ket{\psi},
\end{equation}
where $\ket{\psi}$ is the wavefunction. Since the Hamiltonian is changed slowly, it is convenient to write the wavefunction in the adiabatic (instantaneous) basis,
\begin{equation}\label{SMeq:wav}
    \ket{\psi(t)}=a_+(t)\ket{+(t)}+a_-(t)\ket{-(t)},
\end{equation}
with the eigenstates $\ket{\pm(t)}$ given by~\eqref{eigenS}. The eigenstates and corresponding eigenenergies implicitly depend on time through the coupling parameter. Substituting expansion \eqref{SMeq:wav} into \eqref{SMeq:schro}, and multiplying it by $\bra{+}$ (to shorten the notation we drop the $t$ dependence in the states),
\begin{equation}
        i\partial_t a_+(t)+i a_-(t)\bra{+}\partial_t\ket{-} = E_+(t)a_+(t).
\end{equation}
We now perform a gauge transformation,
\begin{equation}
        a_\pm(t)=\alpha_\pm(t)\exp{\left[-i\Theta_\pm(t)\right]},
\end{equation}
where 
\begin{equation}
    \Theta_\pm(t)=\int_{t_i}^t d\tau E_\pm(\tau).
\end{equation}
The lower integration limit of the previous expression is arbitrary. Under this transformation, the Schrödinger equation becomes
\begin{equation}
    \dot{\alpha}_+(t)=-\alpha_-(t)\bra{+}\partial_t\ket{-}\exp{\left[i(\Theta_+(t)-\Theta_-(t))\right]},
\end{equation}
which can also be rewritten as an integral equation
\begin{equation}
    \alpha_+(t)=-\int_{t_i}^t dt' \alpha_-(t')\bra{+}\partial_{t'}\ket{-}\exp{\left[i(\Theta_+(t')-\Theta_-(t'))\right]}.
\end{equation}
We now compute the first order correction to the wavefunction assuming for simplicity that initially the system is in the pure state $\ket{-}$, so that $\alpha_-(0)=1$ and $\alpha_+(0)=0$. In the slowly-varying parameter limit, we assume that the $\ket{+}$ state will be barely populated, and derive
\begin{equation}
    \alpha_+(t)\approx-\int_{t_i}^t dt'\bra{+}\partial_{t'}\ket{-}\exp{\left[i(\Theta_+(t')-\Theta_-(t'))\right]}.
\end{equation}
The transition probability to the excited state is determined by $\lvert\alpha_+(\tau)\rvert^2$. Using the standard rules for evaluating  the integrals of fast oscillating functions we find
\begin{equation}
    \alpha_+(\tau)\approx\left[i\frac{\bra{+}\partial_t\ket{-}}{E_+(t)-E_-(t)}-\frac{1}{E_+(t)-E_-(t)}\frac{d}{dt}\frac{\bra{+}\partial_t\ket{-}}{E_+(t)-E_-(t)}+...\right]e^{i(\Theta_+(t)-\Theta_-(t))}\Bigg\rvert^{\tau}_{t_i}.
\end{equation}
The first term should dominate in slow processes. In particular, for the Landau-Zener Hamiltonian, this term takes the form
\begin{equation}\label{SMeq:adiabatic}
    \alpha_+(\tau)\approx i\left[\frac{h_x\dot{h}_z(\tau)}{\left(h_x^2+h_z(\tau)^2\right)^{3/2}}e^{i(\Theta_+(\tau)-\Theta_-(\tau))}-\frac{h_x\dot{h}_z(0)}{\left(h_x^2+h_z(0)^2\right)^{3/2}}\right].
\end{equation}
Now, from Eq.\eqref{control}, we also obtain
\begin{equation}\label{SMeq:adiabaticAprox}
    \dot{h}_z(t)=\frac{\left(\dddot{f}_z+\dot{f}_zh_x^2\right)\sqrt{c_1-f_z^2-\left(\frac{\dot{f}_z}{h_x}\right)^2}+\left(\ddot{f}_z+f_zh_x^2\right)\frac{\dot{f}_zf_zh_x^2+\ddot{f}_z\dot{f}_z}{h_x^2\sqrt{c_1-f_z^2-\left(\frac{\dot{f}_z}{h_x}\right)^2}}}{c_1-f_z^2-\left(\frac{\dot{f}_z}{h_x}\right)^2}\frac{1}{h_x},
\end{equation}
but given the boundary conditions $\dot{f}_z(t_B)=\ddot{f}_z(t_B)=0$, 
\begin{equation}
    \dot{h}_z(t_B)=\frac{\dddot{f}_z(t_B)\sqrt{h_x^2+h_z(t_B)^2}}{h_x^2}.
\end{equation}

The auxiliary function $f_z$ is constructed as a function of the dimensionless variable $s=t/\tau$. Consequently, the third derivative of $f_z$ scales as
\begin{equation}\label{dddotf3}
    \dddot f_z(t)=\left(\frac{ds}{dt}\right)^3\frac{d^3}{ds^3}f(s)=\frac{1}{\tau^3}\frac{d^3}{ds^3}f(s).
\end{equation}
Since $h_x$ and $h_z(t_B)$ are unrelated to the protocol duration $\tau$, the first derivative of $h_z$ is also proportional to $1/\tau^3$ at the boundaries, according to Eq.~\eqref{dddotf3}. Thus, when examining Eq.~\eqref{SMeq:adiabaticAprox}, it becomes clear that, for sufficiently long times for the adiabatic approximation to hold in every momentum subspace, the probability of excitation in each of the subspace $\lvert\alpha_+(\tau)\rvert^2$ scales as $\tau^{-6}$. Notably, the boundary conditions for the third derivative of the auxiliary function are entirely arbitrary. Specifically, they can be chosen to be 0. In such a case, the second term in Eq.~\eqref{SMeq:adiabatic} becomes dominant, ultimately leading to an excitation coefficient $\alpha_+(\tau)$ proportional to the fourth derivative of the auxiliary function, thus leading to a scaling of the excitation probability following $\tau^{-8}$. More generally, if $k$ successive derivatives of $f_z$ are engineered to be 0 at the boundaries by choosing a higher order polynomial function with the right coefficients, the probability of excitation should follow a $\tau^{-2k}$ scaling.

\subsection{Minimum evolution time and the quantum speed limit}\label{App:B.2}

Taking polynomial of degree $5$ as an ansatz for $f_z(t)$, we obtain (here using $c_1=1$)
\begin{align}
    f(t)=\frac{1}{\tau^5}\left(-g_0\frac{1}{\sqrt{\Delta^2+g_0^2}}(t-\tau)^3(6t^2+3\tau t +\tau^2)+g_1 \frac{1}{\sqrt{\Delta^2+g_1^2}}t^3(6t^2-15 t\tau+10 \tau^2)\right)
\end{align}
where $h_z(t)\equiv g(t)$, so that $g(t=0)=g_0$ and $g(\tau)=g_1$, and $h_x\equiv \Delta$. From this expression, one finds $g(t)$ using Eq.~\eqref{control}. The expression for the control is a bit involved, 
\begin{align}
    g(t)=\frac{N(t)}{M(t)}
\end{align}
with
\begin{align}
    N(t)=&\left(g_1\frac{t(6t^2(20+\Delta^2t^2)-15t(12+\Delta^2t^2)\tau+10(6+\Delta^2 t^2)\tau^2)}{\sqrt{\Delta^2+g_1^2}}\right.\\&\left.+g_0 \frac{(-6t^3(20+\Delta^2t^2)+15t^2(12+\Delta^2t^2)\tau-10t(6+\Delta^2t^2)\tau^2+\Delta^2\tau^5)}{\sqrt{\Delta^2+g_0^2}}\right)
\end{align}
and 
\begin{align}
    M(t)&=\Delta \left(\tau^5\left[1-\frac{900(g_0(\Delta^2+g_0)^{-1/2}-g_1(\Delta^2+g_1^2)^{-1/2})^2t^4(t-\tau)^4}{\Delta^2\tau^{10}}\right.\right. \\&\left.\left.-\frac{1}{\tau^{10}}\left(g_1(\Delta^2+g_1^2)^{-1/2}t^3(-6t^2+15 t\tau -10 \tau^2)+g_0(\Delta^2+g_0^2)^{-1/2}(t-\tau)^3(6t^2+3t\tau+\tau^2)\right)^2 \right]^{1/2}\right)
\end{align}
Based on the aforementioned ansatz, this method works as long as the term inside the square root of $M(t)$ is non-negative. In order to determine the minimum evolution time, $\tau_{\rm min}$, we consider that $g_1=-g_0$. 
Considering these parameters, the term inside the square root in $M(t)$ becomes negative for $\tau<\tau_{\rm min}$, where
\begin{align}
\tau_{\rm min}=\frac{15g_0}{4\Delta\sqrt{\Delta^2+g_0^2}}
\end{align}
For $\Delta\ll g_0$, this leads to $\tau_{\rm min}\approx 15/(4\Delta)\approx 15/(4\pi)\  \pi/\Delta\approx 1.2\tau_{\rm QSL}$ with $\tau_{\rm QSL}=\pi/\Delta$ the quantum speed limit. Note that this specific value depends on the initial ansatz for $f_z(t)$. This means that $\tau_{\rm min}$ can be further reduced considering other functional forms for $f_z(t)$.

\section{Application to quantum many-body systems}\label{App:C}

\subsection{Transverse-field Ising model}\label{App:C.1}
We consider the Transverse Field Ising Model in one dimension with periodic boundary conditions, this is a chain of $N$ spin-$1/2$ particles separated by a distance $a$ with the ends of the chain connected. The Hamiltonian of such a system is given by
\begin{equation}
    \hat{H}_{\rm TFIM}=-J\sum_{i=1}^N(g \hat{\sigma}_i^x+ \hat{\sigma}_{i}^z\hat{\sigma}_{i+1}^z),
\end{equation}
where $\hat\sigma_i^{x,y,z}$ are the usual Pauli matrices for the spin-$1/2$ particle located at the site $i$, $J$ describes the energy scale of the interaction Hamiltonian, and $g$ is a dimensionless parameter related to the interaction of the spin-$1/2$ particles with an external magnetic field. We assume that $g$ is a controllable parameter. Pauli matrices at two different sites commute. The above system can be mapped to a model of spinless fermions using the Jordan-Wigner transformation, in which a spin up (down) at any site is mapped to the presence (absence) of a spinless fermion at that site. This is done by introducing a fermion annihilation operator $\hat c_i$ such that 
\begin{align}
  \hat \sigma_i^x&=1-2\hat c_i^\dagger \hat c_i,\\
  \hat \sigma_i^z&=-(\hat c_i+\hat c_i^\dagger)\prod_{j<i}(1-2\hat c_j^\dagger \hat c_j),
\end{align}
and $\{\hat c_i,\hat c_j\}=\{\hat c_i^\dagger,\hat c_j^\dagger\}=0$ and $\{\hat c_j,\hat c_i^\dagger\}=\delta_{i,j}$. In this manner, the Hamiltonian can be rewritten as
\begin{equation}
H_{\rm TFIM}=-J\sum_{i=1}^N\left\{ g(1-2\hat c_i^\dagger c_i)+\left[(\hat c_i+\hat c_i^\dagger)\prod_{j<i} (1-2\hat c_j^\dagger \hat c_j)\right] \left[(\hat c_{i+1}+\hat c_{i+1}^\dagger)\prod_{k<i+1}(1-2\hat c_k^\dagger \hat c_k)\right] \right\}.
\end{equation}
Now, using the commutation relations, we find
\begin{align}
  (\hat c_i+\hat c_i^\dagger)\prod_{j<i} (1-2\hat c_j^\dagger \hat c_j)(\hat c_{i+1}+\hat c_{i+1}^\dagger)\prod_{k<i+1}(1-2\hat c_k^\dagger \hat c_k)&= (\hat c_i+\hat c_i^\dagger)(\hat c_{i+1}+\hat c_{i+1}^\dagger)(1-2\hat c_i^\dagger c_i)\\
 &=-(\hat c_{i+1}+\hat c_{i+1}^\dagger)(\hat c_{i}+\hat c_{i}^\dagger)(1-2\hat c_i^\dagger \hat c_i)\\
 &=-(\hat c_{i+1}+\hat c_{i+1}^\dagger)(\hat c_i^\dagger-\hat c_i)=\hat c_{i+1}\hat c_n+\hat c_n^\dagger c_{i+1}+{\rm H.c.},
\end{align}
so that the Hamiltonian reads
\begin{align}\label{eq:HTFIMcn}
\hat H_{\rm TFIM}=-J\sum_{i=1}^N\left(g/2-g \hat c_i^\dagger \hat c_i+\hat c_{i+1}\hat c_i+\hat c_i^\dagger \hat c_{i+1}+{\rm H.c.} \right).
  \end{align}
Note that, by taking $N$ even, the $\hat c_i$'s satisfy either periodic boundary conditions $\hat c_{N+1} = \hat c_1$ (negative parity), or antiperiodic boundary conditions (positive parity), $\hat c_{N+1} = -\hat c_1$. We select the positive parity (as it contains the ground state for $N$ even in the paramagnetic phase). We continue the diagonalization by Fourier transforming these fermionic operators (from now on subscript $k$ or $q$ refers to Fourier transformed, in contrast to $i$ or $j$ for the original operators):
\begin{align}
\hat c_j=\frac{e^{-i\pi/4}}{\sqrt{N}}\sum_k \hat c_k e^{i k j a},  \end{align}
where the sum runs from negative to positive $k$ values ($a$ accounts for the spin inter-spacing). The values that $k$ can take are
\begin{align}
k_j=\pm \frac{(2j-1)\pi}{N a}\  \ j=1,2,\ldots,N/2.
\end{align}
This transformations satisfies $\hat c_{N+1}=-\hat c_1$. In particular, we find the following relations
\begin{align}
  \hat c_j^\dagger \hat c_j&=\frac{1}{N}\sum_{k,q} \hat c_k^\dagger \hat c_q e^{-i j a(k-q)}\\
  \hat c_{j+1}\hat c_j&=\frac{e^{-i\pi/2}}{N}\sum_{k,q} \hat c_k \hat c_q e^{i j a(k+q)}e^{i k a}\\
  \hat c^\dagger_j \hat c_{j+1}&=\frac{1}{N}\sum_{k,q} \hat c_k^\dagger \hat c_q e^{-i j a(k-q)}e^{-i k a}\\
  \hat c_{j}^\dagger \hat c_{j+1}^\dagger &=\frac{e^{i\pi/2}}{N}\sum_{k,q} \hat c_k^\dagger \hat c_q^\dagger e^{-i j a(k+q)}e^{-i k a}\\
  \hat c^\dagger_{j+1} \hat c_{j}&=\frac{1}{N}\sum_{k,q} \hat c_k^\dagger \hat c_q e^{-i j a(k-q)}e^{-i k a}
\end{align}
which upon the sum over all the sites leads to
\begin{align}
    \sum_{j=1}^N  \hat c_j^\dagger \hat c_j&=\frac{1}{N}\sum_{j=1}^N\sum_{k,q} \hat c_j^\dagger \hat c_q e^{-i j a(k-q)}=\sum_k \hat c_k^\dagger \hat c_k,
\end{align}
where we have used $\sum_{j=1}^N e^{-ij a(k-q)}=\delta_{k,q} N$ and the sum over $k$ has been taken for positive and negative values. Similarly,
\begin{align}
  \sum_{j=1}^N \hat c_{j+1}c_j&=e^{-i\pi/2}\sum_{k}\hat c_k \hat c_{-k} e^{i k a}\\
  \sum_{j=1}^N \hat c^\dagger_{j}\hat c_{j+1}&=e^{i\pi/2}\sum_{k}\hat c_k^\dagger \hat c_{-k}^\dagger e^{-i k a}\\
  \sum_{j=1}^N \hat c_{j}^\dagger \hat c_{j+1}&=\sum_{k}\hat c_k^\dagger \hat c_{k} e^{-i k a}\\
    \sum_{j=1}^N c_{j+1}^\dagger c_{j}&=\sum_{k}\hat c_k^\dagger \hat c_{k} e^{i k a}.
\end{align}
Introducing the previous expressions into Eq.~\eqref{eq:HTFIMcn}, we obtain
\begin{align}
  \hat H_{\rm TFIM}&=J\sum_k\left[-g +2g \hat c_k^\dagger \hat c_k-e^{-i\pi/2}\hat c_k\hat c_{-k} e^{ika}-\hat c_k^\dagger \hat c_k e^{ika}-e^{i\pi/2}\hat c_{-k}^\dagger \hat c_k^\dagger e^{-i ka}-\hat c_k^{\dagger}\hat c_k e^{-ika} \right]\\
  &=J\sum_k \left[-g +(2g-e^{ika}-e^{-ika})\hat c_k^\dagger \hat c_k-e^{-i\pi/2}\hat c_k \hat c_{-k}e^{ika}-e^{i\pi/2}\hat c^\dagger_{-k}\hat c_k^\dagger e^{-ika} \right].
\end{align}
Now using
\begin{align}
  \sum_k \left(i \hat c_k \hat c_{-k} e^{i k a}-i \hat c_{-k}^\dagger \hat c_k^\dagger e^{-i ka}\right)&=\sum_{k>0} \left( i e^{i ka} \hat c_k \hat c_{-k}-i e^{-ika}\hat c^\dagger_{-k} \hat c_k^\dagger+i e^{-i k a}\hat c_{-k}\hat c_k-ie^{i ka} \hat c_k^\dagger \hat c_{-k}^\dagger\right)\\&=\sum_{k>0} (-2) \sin ka \left(\hat c_k\hat c_{-k}+\hat c_{-k}^\dagger \hat c_k^\dagger\right)=-\sum_k \sin ka\left( \hat c_k \hat c_{-k}+\hat c_{-k}^\dagger \hat c_k^\dagger\right),
\end{align}
we finally obtain
\begin{align}
\hat H_{\rm TFIM}&=J\sum_k\left[-g+2(g-\cos ka)\hat c_k^\dagger \hat c_k-\sin ka\left( \hat c_k \hat c_{-k}+\hat c_{-k}^\dagger \hat c_k^\dagger\right) \right].
\end{align}
Let us now define a fermionic mode $\Psi^\dagger_k=(\hat c_k^\dagger,\hat c_{-k})$, so that we can write
\begin{align}
    \hat H_{\rm TFIM}=\sum_{k>0} \Psi^\dagger_k \hat H_k \Psi_k
\end{align}
where $\hat H_k$ is given by
\begin{align}
\hat H_k=h_k^z(g) \frac{\hat \sigma^k_z}{2}+h_k^x \frac{\hat \sigma^k_x}{2},
\end{align}
with $\hat \sigma^k_z=\ket{1}_k\bra{1}_k-\ket{0}_k\bra{0}_k$, 
$h_k^z(g)=4J(g-\cos (k a))$, and $h_k^x=4J\sin (k a)$. 

The control of $g(t)$ is designed to follow the eigenstates of the invariant associated with the lowest-energy Hamiltonian, i.e., 
\begin{equation}
\hat H_{k_0} = 2J\left[ g - \cos\left(\frac{\pi}{N}\right)\right]\hat\sigma_{z}^{k_0} +2J\sin\left(\frac{\pi}{N}\right) \hat\sigma_{x}^{k_0}.
\end{equation}
According to Eq.~\eqref{control}, $g(t)$ is defined as
\begin{equation}
    g(t) =\cos\left(\frac{\pi}{N}\right)+\frac{1}{4J}\frac{\ddot f_z+f_z\left(4J\sin\left(\frac{\pi}{N}\right)\right)^2}{\sqrt{K-f_z^2-\frac{\dot f_z^2}{\left(4J\sin\left(\frac{\pi}{N}\right)\right)^2}}}\frac{1}{\left(4J\sin\left(\frac{\pi}{N}\right)\right)},
\end{equation}
with $f_z=f_z(t)$ any arbitrary function that fulfills the boundary conditions given by Eq.~\eqref{co1}.

For comparison, we also design the control according to the FAQUAD method. The adiabaticity parameter for the lowest-energy subspace is, according to Eq.~\eqref{adia},
\begin{equation}\label{smeq:faquad}
    \mu = \frac{\dot{g}(t)\sin\left(\frac{\pi}{N}\right)}{8J\left[g^2(t)+1-2g(t)\cos\left(\frac{\pi}{N}\right)\right]^{3/2}}.
\end{equation}
Equation~(\ref{smeq:faquad}) can be integrated considering $\mu={\rm const}$. The value of the $\mu$ and the integration constant can be found from the initial conditions $g(0)=g_0$ and $g(\tau)=g_1$. 

As detailed in Section B of the Supplementary Material, it is observed that the FAQUAD method does not yield unit fidelity even within the lowest-energy subspace. Instead, fidelity experiences oscillations as a function of the protocol duration. These oscillations depend on the specific parameters characterizing each subspace, thereby leading to dephased oscillations across the system's subspaces.
Consequently, when employing the FAQUAD approach to control the entire system, achieving high fidelity becomes increasingly challenging as 
the system size grows.

\subsection{Long-range Kitaev chain}\label{App:C.2}
Let us now consider the case of the long-range Kitaev chain of fermionic particles with periodic boundary conditions. The Hamiltonian reads
\begin{equation}
     \hat{H}_{\rm LRK}=-J\sum_{i=1}^N\left[ \sum_{r>0} \left(J_r\left( \hat{c}_i^\dagger \hat{c}_{i+r}+\hat{c}_{i+r}^\dagger \hat{c}_{i}\right)+d_r \left(\hat{c}_i\hat{c}_{i+r}+\hat{c}_{i+r}\hat{c}_{i}\right)\right)-g(t) \hat{n}_i \right],
\end{equation}
where $\hat c_i$ is the fermionic annihilation operator at site $i$, $\hat n_i=\hat c_i^\dagger\hat c_i$, $g(t)$ is a dimensionless controllable parameter related to the chemical potential, and
\begin{equation}
    J_r=\frac{1}{N\bar{r}^\alpha},\hspace{6mm}d_r = \frac{1}{N\bar{r}^\beta},
\end{equation}
where, given the periodic boundary conditions, $\bar{r}=a{\rm min}(r,N/2-r)$. The Hamiltonian is quadratic in fermions and can hence be exactly solved by a Fourier Transformation in the momentum space $\hat c_i=\frac{1}{\sqrt{N}}\sum_{n=0}^{N-1}e^{-ik_nx_i}\hat{c}_{k_n}$. Assuming anti-periodic boundary conditions $(\hat c_i = -\hat c_{i+N})$, the momenta modes are quantized as $k_n = (2\pi/Na)(n-1/2)$. In this basis, the Hamiltonian can be written as 
\begin{equation}
\hat H_{\rm LRK}=\sum_{k>0}\Psi^\dagger_{k}\hat H_k\Psi_k,
\end{equation}
where the sub-index in $k_n$ has been droped for simplicity, $\Psi^\dagger=(\hat c_k^\dagger,\hat c_{-k})$ and $\hat H_k = h_{z,k}\hat\sigma_z^k/2+h_{x,k}\hat\sigma_x^k/2$. Each subspace of momentum $k$ evolves independently, with $h_{z,k}=g-4J\sum_{r>0}J_r\cos\left(kra\right)$ and $h_{x,k}=-2J\sum_{r>0}d_r\sin(kra)$. 

Just as in the application to the TFIM, the invariant protocol is designed to drive the system with the smallest energy gap ($k_0$) with unit fidelity.

\subsection{Details on the scaling analysis}
\begin{figure}[t!]
\begin{center}
\includegraphics[width=0.7\linewidth]{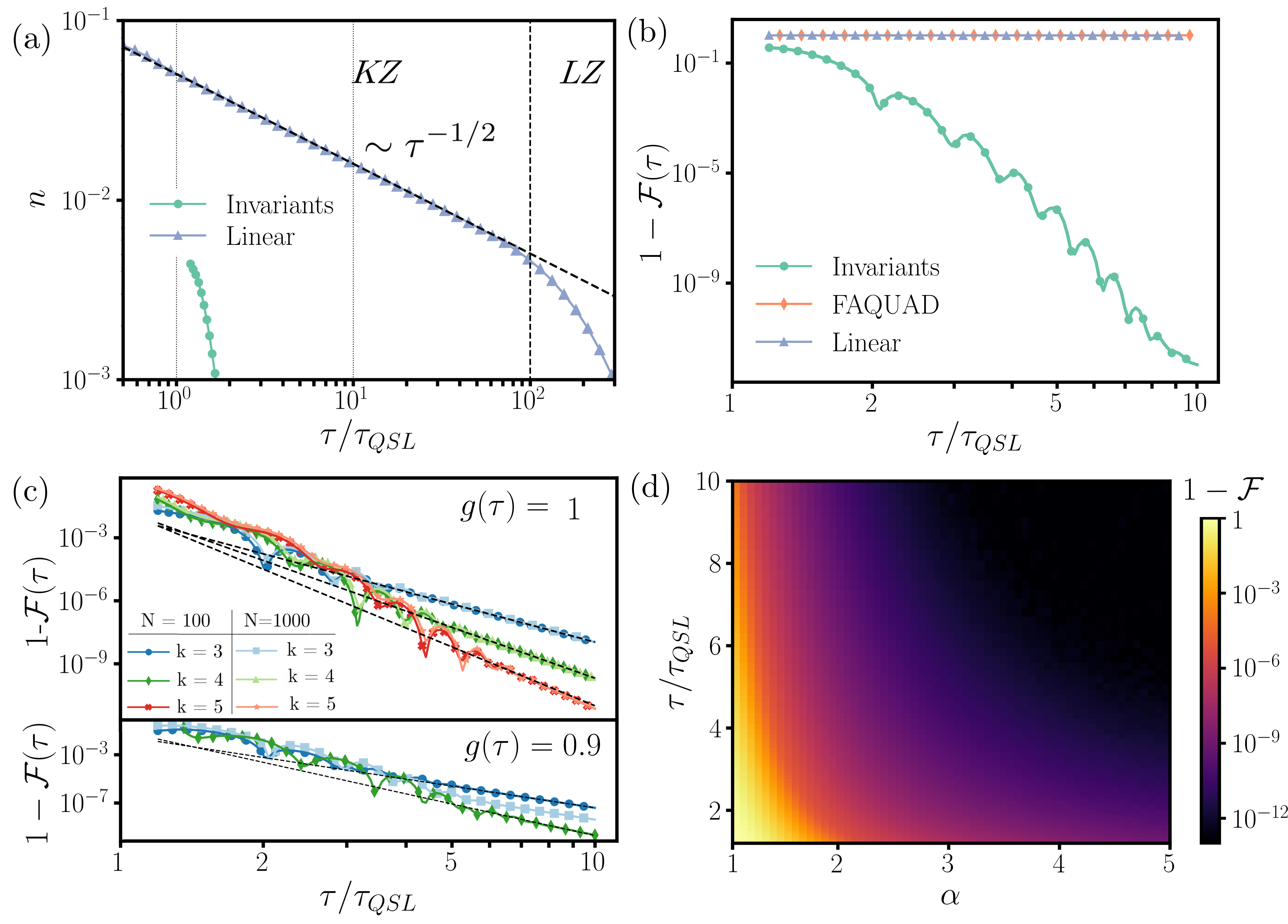}
\caption{(a) Scaling of the TFIM for the linear and invariants protocol. Same parameters as in Fig.\ref{fig1}(b), but using a different scale. The thick dashed line represents the scaling $n\propto\tau^{-1/2}$, while the vertical lines delimit the studied range in Fig.~\ref{fig1}(b) (thin dotted lines), and the deviation from Kibble-Zurek mechanism (thin dashed line). (b) Infidelity for fixed $N=200$ spins as a function of the rescaled quench time $\tau/\tau_{QSL}$  (log-log scale) for the TFIM and the three protocols. Parameter values: $\lambda=1$, $J=1$, $g(0)=10$, $g(\tau)=0$. (c) Scaling of the infidelity for the TFIM invariant protocol for $N=100$ and $N=1000$ as a function of the rescaled quench time for protocols with $k=3$, $k=4$ and $k=5$. The final value of the control parameter is $g(\tau)=1$ (top) and $g(\tau)=0.9$ (bottom). The polynomial behavior cannot be observed for $g(\tau)=0.9$ in the cases $k=4$ $N=1000$ or $k=5$  due to limitations in numerical precision. (d) Scaling of the infidelity in the LRK for $N=100$ spins as a function of the rescaled quench time and the long-range exponent $\alpha$. Parameter values: $\beta \rightarrow \infty$, $J=2$, $g(0)=10$, $g(\tau)=0$.}
\label{Infidelity_supp}
\end{center}
\end{figure}

In the main text, we have demonstrated the scaling behavior of the density of excitations, denoted as $n$, which is defined as $n=1-\frac{2}{N}\sum_{k>0}|\langle \phi_{k,0}|\psi_k(\tau)\rangle|^2$. This scaling has been examined with respect to two key parameters: the rescaled quench time, $\tau/\tau_{QSL}$, and the total number of interacting particles, denoted as $N$. In Fig.\ref{Infidelity_supp}, we present a detailed analysis of the density of excitations produced by the linear protocol across an extended range of driving times. Remarkably, we observe that the density of excitations follows a scaling behavior of $\tau^{-1/2}$ within the range $\tau\lesssim 100\tau_{QSL}$, consistent with the predictions of the Kibble-Zurek mechanism. This compelling observation underscores the significant breakdown of the expected Kibble-Zurek scaling behavior induced by the protocol based on dynamical invariants.

For the sake of completeness and a comprehensive understanding of the system's behavior, we also present an analysis of the scaling properties of the infidelity, denoted as $\mathcal{I}$, which is defined as $\mathcal{I}=1-\mathcal{F}$. Here, $\mathcal{F}$ represents the fidelity and is defined as $\mathcal{F}=\langle\phi_0(g_1)|\psi(\tau)\rangle|^2$. We perform this analysis across the same set of parameters, namely $\tau/\tau_{QSL}$ and $N$, in Fig.\ref{Infidelity_supp}(b-d). This examination serves to verify the equivalence between the scaling behaviors of these two fundamental measures, providing a more comprehensive perspective on the quantum phase transitions under investigation.
%
%
%
%

\subsection{Details on robustness}\label{App:C.4}
The robustness of the invariant-based control is studied in the TFIM. The Hamiltonian of interest can be written as
\begin{align}
\hat{H}_{\rm TFIM}=-J\left(g(t)\sum_{i=1}^N\hat{\sigma}_i^x+ \sum_{i}^N\lambda_i \hat{\sigma}_i^z\hat{\sigma}_{i+1}^z\right).
\end{align}
Note that we have now included site-dependent interaction strength $\lambda_i$. In addition, as mentioned in the main text, we analyze the impact on noisy controls, namely, $g(t)\rightarrow g(t)+\eta(t)$ so that $\eta(t)$ represents a standard white Gaussian noise, $\langle \eta(t)\rangle=0$ and $\langle \eta(t)\eta(t')\rangle=W^2$ being $W$ its strength.  The impact of these two different imperfections is studied individually.

\subsubsection{Noisy control}
First, the noisy control can be tackled following similar steps as in Ref.~\cite{Dutta:16}. For this, we take $\lambda_i=1 \ \forall i$, i.e. no interaction disorder. The wavefunction evolves according to
\begin{align}
    \frac{d}{dt}\ket{\psi(t)}=-i(\hat{H}_{0}(t)+\eta(t) \hat{H}_1)\ket{\psi(t)}
\end{align}
where $\hat{H}_0(t)$ is the sure and ideal TFIM Hamiltonian, while $\hat{H}_1=-J\sum_{i=1}^N \hat{\sigma}_i^x$.  Averaging over the noise the state is described in terms of a density matrix $\hat{\rho}(t)$ which obeys the master equation
\begin{align}
    \frac{d}{dt}\hat{\rho}(t)=-i[\hat{H}_0(t),\hat{\rho}(t)]-\frac{W^2}{2}\left[\hat{H}_1,[\hat{H}_1,\hat{\rho}(t)]\right].
    \end{align}
Upon Jordan-Wigner and Fourier transformations, and assuming a product state in the momentum space, we arrive at the master equation for the density matrix for the $k$ subspace
\begin{align}
    \frac{d}{dt}\hat{\rho}_k(t)=-i[\hat{H}_k(t),\hat{\rho}_k(t)]-\frac{W^2}{2}[\hat{H}_{1,k},[\hat{H}_{1,k},\hat{\rho}_k(t)]],
\end{align}
with $\hat{H}_{1,k}=2J\hat{\sigma}_{z}^k$, while $\hat{H}_k(t)$ has the same form as above. The results for different values of $W$ are plotted in the main text (cf. Fig. 2). While the previous master equation captures the leading order effect of the noisy control, it is important to remark that terms mixing momentum subspaces have been discarded as they play no role in determine single-particle observables such as $n$ (see~\cite{Dutta:16,Nalbach:15}).

\subsubsection{Random disorder}
Second, we examine the impact of interaction-strength disorder. For that, the dimensionless coupling strength $\lambda_i$ is considered to be a uniform random variable such that $\langle \lambda_i\rangle = 1$ and $\lambda_i\in[1-\Lambda,1+\Lambda]$, and independent among sites.  Note that one could consider a global noisy coupling strength $\lambda\rightarrow \lambda +\eta(t)$ for all sites, whose impact would be similar to a noisy control. This interaction-strength disorder is tackled by relying on a Jordan-Wigner transformation of the TFIM. The fermionic Hamiltonian results in
\begin{align}
    \hat{H}_{\rm TFIM}=-J\left[\sum_{i=1}^{N-1}\lambda_i(c_i^\dagger c_{i+1}^\dagger+c_i^\dagger c_{i+1}+{\rm H.c})-2g\sum_{i=1}^N c_i^\dagger c_i\right]+J\lambda_N (c_N^\dagger c_1^\dagger+c_N^\dagger c_1+{\rm H.c.}),
\end{align}
since we consider periodic boundary conditions. Since the couplings are not homogeneous, this model cannot be brought into a collection of independent Landau-Zener or two-level systems in the momentum space. Instead, one needs to solve this fermionic system, which can be written as 
\begin{align}
\hat{H}_{\rm TFIM}=\Psi^\dagger \hat{H}_f \Psi\end{align}
with $\Psi^\dagger=(\hat{c}_1^\dagger, \hat{c}_2^\dagger,\ldots,\hat{c}_N^\dagger,\hat{c}_1,\hat{c}_2,\ldots,\hat{c}_N)$ which fulfill the fermionic anti-commutation relations $\{ \Psi_i,\Psi_j^\dagger\}=\delta_{i,j}$, and 
\begin{align}\label{eq:Hf}\hat{H}_f=\begin{bmatrix} A& B \\ -B & -A\end{bmatrix}\end{align} with
$A$ and $B$ symmetric and anti-symmetric real matrices with matrix-elements $A_{i,i}=Jg$, $A_{i,i+1}=A_{i+1,i}=-J\lambda_i/2$, but $A_{N,1}=A_{1,N}=J\lambda_N/2$, while $B_{i,i+1}=-B_{i+1,i}=-J\lambda_i/2$, and $B_{1,N}=-B_{N,1}=J\lambda_N/2$. 
In the Heisenberg picture, and employing a Bogoliubov operator $c_{i,H}(t)=\sum_{\mu=1}^N(u_{i,\mu}(t)\gamma_\mu +v_{i,\mu}^*(t)\gamma_\mu^\dagger)$ with fermionic operators $\gamma_\mu$, it can be shown that the coherent dynamics is fully characterized by the evolution of the coefficients 
\begin{align}
    \dot{u}_{i,\mu}&=-2i\sum_{j=1}^N\left[A_{i,j}(t)u_{j,\mu}(t)+B_{i,j}v_{j,\mu}(t)\right]\\
    \dot{v}_{i,\mu}&=2i\sum_{j=1}^N\left[A_{i,j}(t)v_{j,\mu}(t)+B_{i,j}u_{j,\mu}(t)\right]
\end{align}
with the initial condition set at by the eigenvalue problem in Eq.~\eqref{eq:Hf} at $g(t=0)=g_0$. Note that the time-dependent control enters via the symmetric matrix $A(t)$, and $u(t)$ and $v(t)$ are $N\times N$ matrices that satisfy  $u^\dagger(t) u(t) +v^\dagger(t) v(t)=\mathbb{I}$ at all times to ensure the fermionic anti-commutation relation.

We then numerically solve the coupled differential equations under the protocol $g(t)$ for $u_{i\mu}(t)$ and $v_{i\mu}(t)$ for different disorder realizations and then average the results. We compute the density of excitations, which in the original spin representation reads as
\begin{align}
n_d=\frac{1}{N}\sum_{i=1}^N\bra{\psi(\tau)}\frac{1}{2}(1-\hat{\sigma}^z_i\hat{\sigma}^z_{i+1})\ket{\psi(\tau)}.
\end{align}
In the fermionic Jordan-Wigner representation, it results in $n_d=\frac{1}{2N}\sum_{i=1}^{N-1}(1-[(v(\tau)-u(\tau))(u^\dagger(\tau)+v^\dagger(\tau))]_{i+1,i})+\frac{1}{2N}(1-[(v(\tau)-u(\tau))(u^\dagger(\tau)+v^\dagger(\tau))]_{1,N})$.  

Recall that the invariant-based protocol is not altered, i.e. we perform the protocol designed for an ideal TFIM (without the disorder, $\Lambda=0$).  The average density of excitations is plotted in the main text (cf. Fig. 2).

\bibliography{paper.bib}

\end{document}